\documentclass[]{aa}
\usepackage{txfonts}
\begin{document}

\title{Binary corrected X-ray light curve of Cygnus X-3 : implications 
for the timing properties of the compact binary system}
\author{M.~Choudhury\inst{1}, A.~R.~Rao\inst{1}, S.~V.~Vadawale\inst{1}\fnmsep\thanks{\emph{Present address:} Harvard Smithsonian Center for Astrophysics, 60 Garden Street, Cambridge, MA 02138}, A.~K.~Jain\inst{2} \& N.~S.~Singh\inst{3}}
\offprints{M. Choudhury \email{manojulu@tifr.res.in}} 
\institute{Tata Institute of Fundamental Research, Homi Bhabha Road, Mumbai-400005, India
\and ISRO Satellite Centre, Bangalore-560017, India
\and Dept. of Physics, Manipur University, Canchipur, Imphal, Manipur-795003, India.
	}
\date{Received 2 May 2003 / Accepted 9 March 2004}
\authorrunning{Choudhury et al.}
\titlerunning{Binary corrected X-ray light curve of Cygnus X-3 }

\abstract{
We study the binary corrected X-ray light curve of Cygnus X-3 using the RXTE-ASM
(All Sky Monitor) data and RXTE-PCA data. We find that the X-ray template derived
from the EXOSAT observations adequately explains the binary variations. Using the
recently determined quadratic binary ephemeris we correct the RXTE-PCA lightcurve and
obtain the power spectrum which shows distinct features of shifting towards the low
frequency regime vis-a-vis the Galactic X-ray binaries. The power  spectrum
doesn't vary from the hard to the soft state. We also  obtain the binary
corrected RXTE-ASM monitoring lightcurve and examine the previously obtained
correlation between soft X-ray (2 -- 12 keV from RXTE ASM), hard X-ray (20 -- 100 keV
from CGRO -- BATSE) and radio (2.2 G Hz from GBI) after the binary correction. We
find that the correlation time scale is less than a day.
}

\maketitle

\section {Introduction}

Cygnus X-3, first discovered as a bright X-ray source (Giacconi et al.\cite{gia67}),
is an extensively studied Galactic binary system, being one of the brightest source
in the X-ray, infra-red and radio regions of the electromagnetic spectrum (see
Bonnet-Bidaud \& Chardin \cite{bon88} and the references therein). It shows a strong
4.8 hour modulation in the X-ray (Parsignault et al. \cite{par72}) and the infra-red
emission which is attributed to the binary orbital motion (Becklin et al.
\cite{bec73}). The measurement of the radial velocity in the infra-red band and its
interpretation as due to binary Doppler shift has led to the derivation of a large
mass function for the source (Schmutz et al. \cite{sch96}), but this simplistic
explanation of the He II line shift is strongly disputed by the relative phasing of
the infra-red and the X-ray binary modulation (van Kerkwijk \cite{van93}). If the
spectral  features observed in the infrared are ascribed to a mass losing Wolf-Rayet
like companion, a large mass loss, large mass for the companion and
$\sim$20 M$_\odot$ mass for the primary (suggesting Cygnus X-3 as a black hole
source) can be derived (van Kerkwijk  et al. \cite{van92}). Mitra (\cite{mit96}),
however, has pointed out that these derived parameters are incompatible with observed
X-ray properties of the source. Interestingly, very recently Stark \& Saia
(\cite{sta03}) have suggested an upper limit to the mass of the compact object of
3.6 M$_{\odot}$. They measured the Doppler shift of the He $\alpha$ like line of Fe
XXV (Paerels et al. \cite{pae00}, Kitamoto et al. \cite{kit94}) and ascribing it to
originate from very near the surface of the compact object they provide a
conservative estimate of the stellar masses and the separation of the binary system. 

The 4.8 hour X-ray light curve has shown a remarkable stability in terms of its shape
(Ghosh et al. \cite{gho81}) and the period has shown a gradual increase at a rate of
$\sim$10$^{-9}$ s s$^{-1}$. Recently, Singh et al. (\cite{sin02}) have extended the
span of the observation and have concluded that the period change is consistent with
a constant rate of 5.76$\pm$0.24 $10^{-10}$. Remarkably, the binary modulation was
also found to be consistent with the template derived from the EXOSAT data (van der
Klis \& Bonnet-Bidaud \cite{van89}).

The radio emission shows flaring activity (Gregory et al. \cite{gre72}; Watanabe
et al. \cite{wat94}; McCollough et al. \cite{mcc99}) during which it exhibits
radio-jets analogous to those of powerful radio galaxies and quasars (Waltman et al.
\cite{wal95}; Fender et al.\cite{fen97}). There have been several attempts to
connect the radio, X-ray, and infrared variations in the source (Watanabe et al.
\cite{wat94}; Fender \cite{fen97}). Recently Choudhury et al. (\cite{cho02a}), have
reported a strong and significant correlation between the radio and X-ray emission in
the low-hard state of Cygnus X-3. They concluded that the radio emission is related
to the shape of the X-ray  spectrum which pivots at around 12 keV with the X-ray
photons softer than the pivot energy highly correlated to the radio emission. This
signifies a definite connection between the inflow of matter via accretion disk
emission and the outflow via jet emission (Choudhury et al. \cite{cho03}).

In this paper we develop a method to correct for the variation in the X-ray emission
due to the binary modulation in this source and thereafter study the X-ray timing
characteristics. We also investigate the above mentioned radio:X-ray correlation of
the source (Choudhury et al. \cite{cho02a}) after correcting for the binary
modulation in the X-ray and conclude that the correlation time scale is shorter than
one day.

\begin{figure}[t]
\includegraphics{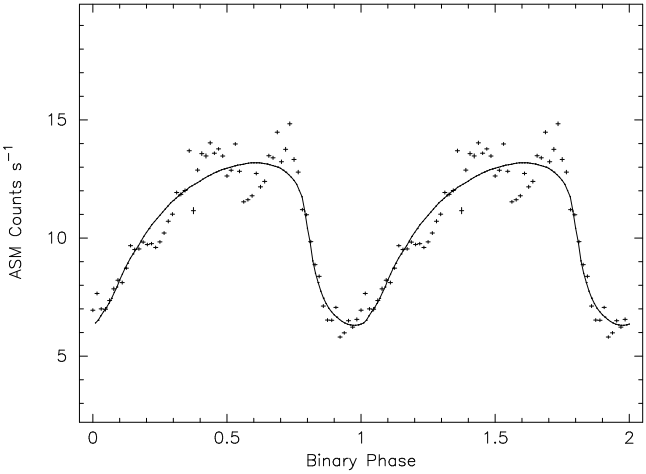}
%\resizebox{\hsize}{!}{\includegraphics[]{grs1915_lc.ps}}
\vskip 6.2cm
\caption{
The X-ray light curve of Cygnus X-3 obtained from RXTE-ASM for the period of MJD
50410 to 51400 folded at the quadratic ephemeris given by Singh et al.
(\cite{sin02}). The template data obtained from EXOSAT (van der Klis \& Bonnet-Bidaud
\cite{van89}) is also shown, with appropriate scaling for the count rates.
}
\label{fig1}
\end{figure}

\section {Binary correction}
The evolution of the 4.8 hour binary period of Cygnus X-3 has been studied
extensively with the time derivative ($\dot{P}$) measured to be $\sim$ 10$^{-9}$
(van der Klis \& Bonnet-Bidaud \cite{van81,van89}; Kitamoto et al. \cite{kit87}).
Recently, Singh et al. (\cite{sin02}) have extended the data base of the binary
period measurements using the Indian X-ray Astronomy Experiment (IXAE) and archival
data from ROSAT, ASCA, BeppoSAX and RXTE. They found that the binary template
obtained from the EXOSAT data adequately explains the recent observations and have
derived a value of the period derivative of 5.76$\pm$0.24 10$^{-10}$. Inclusion of
second derivative marginally improved the fit, essentially giving an upper limit to
the second derivative ($\ddot{P}$ = -1.3 $\pm$1.6 10$^{-11}$ yr$^{-1}$).

\begin{figure}[t]
\includegraphics{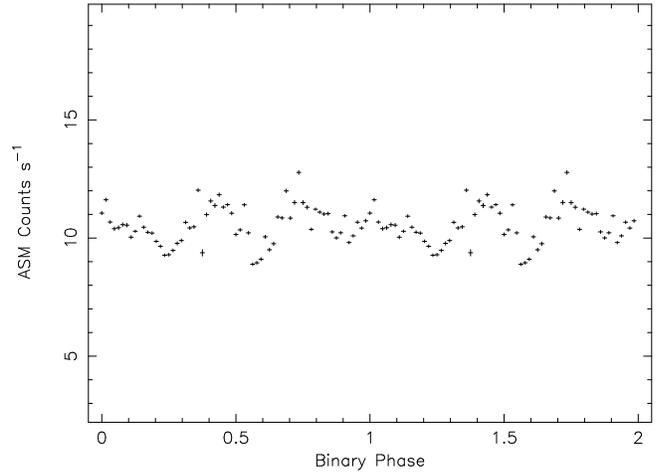}
%\resizebox{\hsize}{!}{\includegraphics[]{grs1915_lc.ps}}
\vskip 6.2cm
\caption{The X-ray light curve of Cygnus X-3, obtained from RXTE-ASM monitoring data,
folded at a binary period of 4.8 hours after the correction for the quadratic
ephemeris given by Singh et al. (\cite{sin02}).
}
\label{fig2}
\end{figure}

We have folded the RXTE ASM data using the quadratic ephemeris of Singh et al.
(\cite{sin02}) from MJD 50410 to MJD 54100 (P = 0.19968443 d;
$\dot{P}$ = 5.76 10$^{-10}$; time of zero phase T$_0$= 2440949.892 JD). The time span
chosen for the folding corresponds to the data used by Choudhury et al.
(\cite{cho02a}) for the detailed X-ray, radio and hard X-ray correlation analysis.
The folded light curve is shown in Figure \ref{fig1} (the errors in the data points
are smaller than the symbol size). The template data from van der Klis \&
Bonnet-Bidaud (\cite{van89})  is also shown in Figure \ref{fig1}, after appropriate
scaling. The relative phase of the data and template are {\it not arbitrarily
shifted} for a proper match: the folded light curve (with the above ephemeris) is
overlaid with the vertically shifted and scaled  template data. The folded count rate
correlates with the template value with a correlation coefficient of 0.95 (for 64
data points). It can be seen from the figure that there are non-statistical
variations due to the source over and above the binary variation. The ASM count rate,
C, can be expressed as 
%\begin{center}
\begin{equation}
\label{CR}
C = 10.50\pm0.59 + 3.31\pm0.35 V
\end{equation}
%\end{center}
where V is the template value for a given phase as given in van der Klis \&
Bonnet-Bidaud (\cite{van89}). The errors are nominal 1 $\sigma$ errors obtained by
assuming that the fluctuations in each phase bin is random in nature.

During this period, Cygnus X-3 shows several X-ray flares and also transition between
low-hard to high-soft states (see Figure \ref{fig1} of Choudhury et al.
\cite{cho02a}). Considering the large variation of the X-ray flux during the period
covered, the folded light curve agrees with the template quite well. We have examined
the fluctuations in the source counts over and above the binary variations with
respect to binary phase as well as the observation duration. Though there is some
evidence for large flare like variations during the phase 0.2 to 0.5, the rest of the
fluctuations occur throughout without any association with binary phase or time of
observations. Note also that we have not done any phase fitting for this analysis.
Hence we conclude that the error in the phase of light curve minimum is negligible
compared to the overall source fluctuations.

Using this information, we can correct each observation for the binary modulation, as
follows. The quadratic ephemeris to get the zero phase for the $nth$ cycle of the
period is given as (Singh et al. \cite{sin02}):
%\begin{center}
\begin{equation}
\label{PE}
T_n = T_0 + P_0 n + c n^2
\end{equation}
%\end{center}
where c = $ { 1 \over 2} P_0 \dot{P}$. This equation can be inverted to
give the binary phase at any time T as,
%\begin{center}
\begin{equation}
\label{PT}
n = { {(T - T_0)} \over P_0} - { {(T - T_0) c }  \over  P_0^3}
\end{equation}
%\end{center}
For the RXTE ASM data, we have calculated the phase using equation \ref{PT} and from
the corresponding template value, corrected for the linear term in equation \ref{CR}.
The binary corrected ASM data is again folded and the folded light curve is shown in
Figure \ref{fig2}. It can be seen from the figure that for the long term monitoring
data a quadratic ephemeris can explain the modulation quite well and this ephemeris
may be used to correct the data for the binary variation. However, we caution, that
despite the stable profile with a monotonic change in its period for several years,
Cygnus X-3 may show jitters in the individual binary phase measurements (see Singh et
al. \cite{sin02}). This implies that either (a) there is an inherent jitter in the
minimum of the binary phase or b) it has a stable profile and short term variability
is superimposed on it. We have made the binary correction for the subsequent analysis
assuming that Cygnus X-3 has a stable profile. The implication of the possible random
phase jitter is examined at appropriate places, if required.

\begin{figure}[t]
\includegraphics{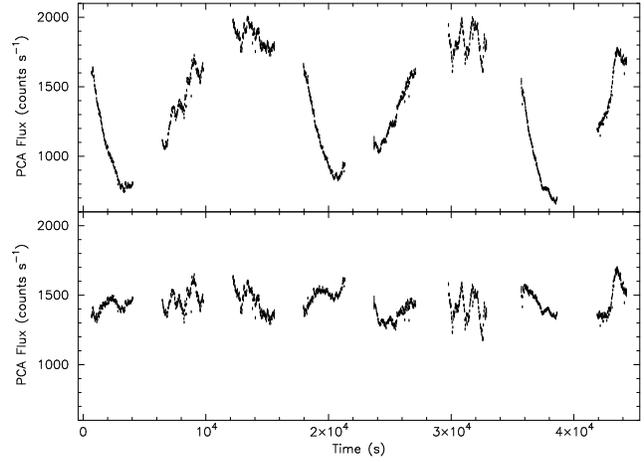}
\vskip 6.2cm
\caption{The X-ray light curve of Cygnus X-3  obtained from RXTE-PCA pointed 
observations (top panel), shown along with light curve (bottom panel)
corrected for the binary variation using the
quadratic ephemeris given by Singh et al. (\cite{sin02}).
}
\label{fig3}
\end{figure}

We apply the binary correction to the individual RXTE-PCA observations and in Figure
\ref{fig3} we show both the uncorrected (top panel) and corrected (bottom panel)
lightcurve of the single longest observation (Obs. Id. 10126-01-01-020 \&
10126-01-01-02) of the source by RXTE. The observation covers about two and a half
binary cycles, punctuated by the necessary breaks due to Earth occultation and
various other data dropouts. The lightcurve is obtained from single bit data (2--6.5 
keV) with all the five PCUs on. The binary template (Figure \ref{fig1}) is asymmetric
with an unusually broad peak lasting in the $\sim 0.4-0.75$ phase of the binary
period with a gradual rise before and a steep fall after the peak. It is evident that
the correction for binary modulation is very good for the rising and falling phase of
the binary ephemeris, highlighting the small variations which were otherwise
smothered by the binary modulation. Again we emphasize here that we have not done any
fitting for deriving the zero phase, but have used the quadratic ephemeris to derive
it. During  the peak the lightcurve shows fluctuations not correctable by the smooth
peak of the template. This, generally random, fluctuation is an inherent feature of
the source present in all the observations, past and present (see van der Klis \&
Bonnet-Bidaud \cite{van82,van89}). We employ this binary correction for the timing
analysis presented in the following sections.

\section{Temporal characteristics of pointed RXTE observations}

\begin{figure}[t]
\includegraphics{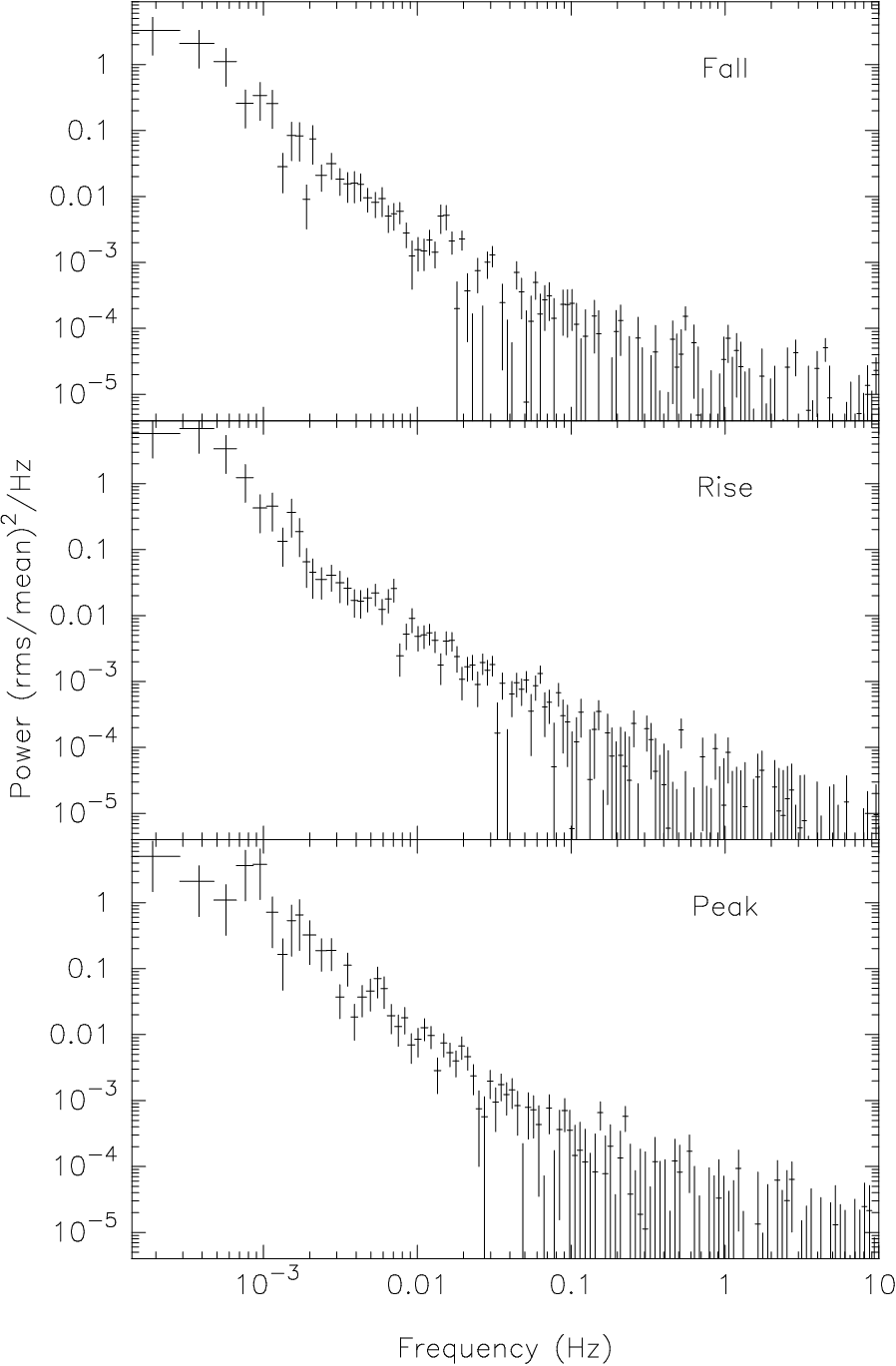}
%\resizebox{\hsize}{!}{\includegraphics[]{grs1915_lc.ps}}
\vskip 8.2cm
\caption{
Power Density Spectrum (PDS) of Cygnus X-3 for the RXTE-PCA pointed observations
shown in Figure \ref{fig3}, after correcting for the binary variations. The PDS is
generated separately for three regions of the binary phase namely
the fall (phase 0.75 - 1.0 - top panel), rise (phase 0 - 0.4 - middle 
panel) and the peak (phase 0.4 - 0.75 - bottom panel).
}
\label{fig4}
\end{figure}

\begin{figure}[t]
\includegraphics{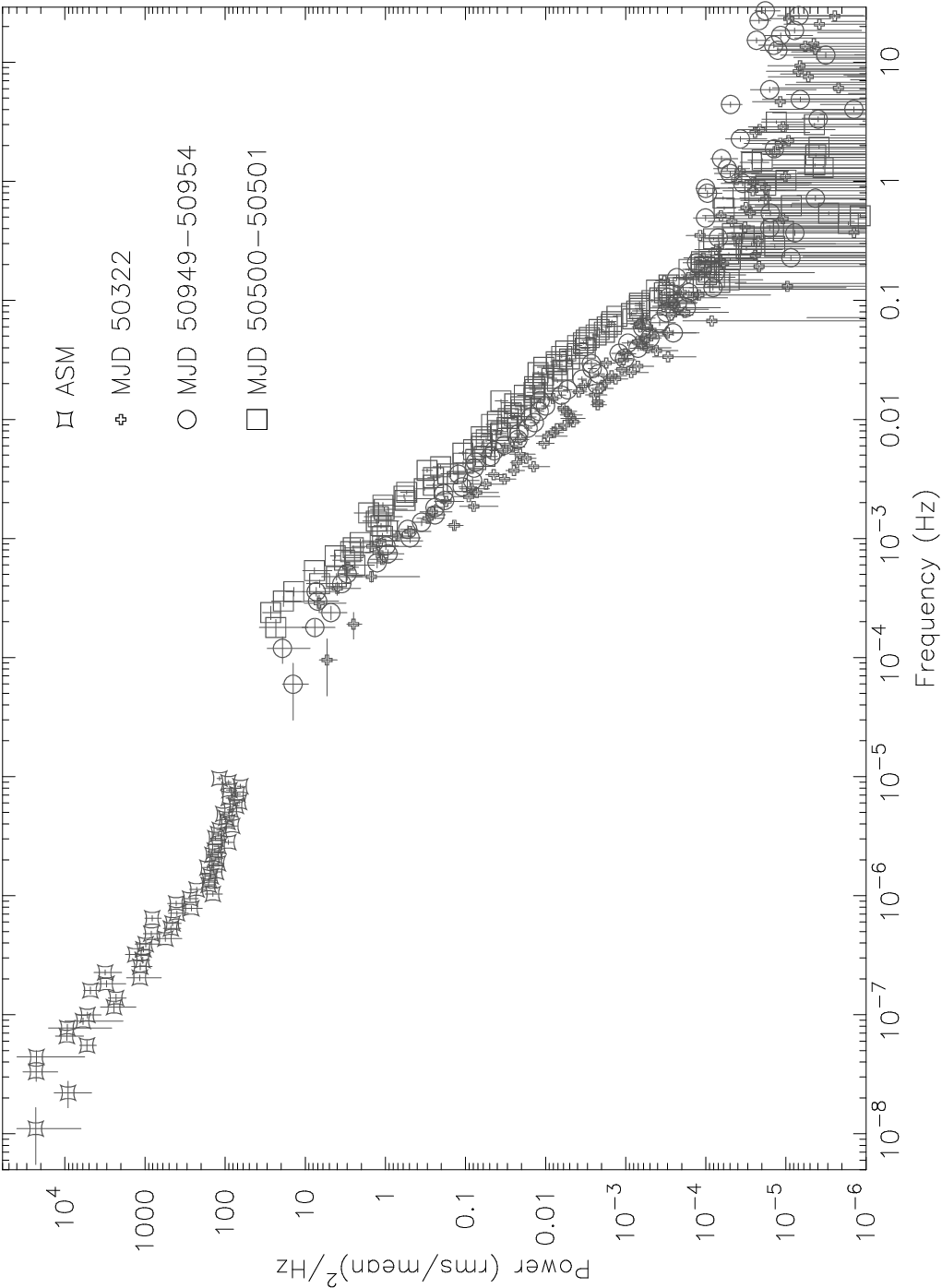}
%\resizebox{\hsize}{!}{\includegraphics[]{grs1915_lc.ps}}
\vskip 7.5cm
\caption{
The PDS of Cygnus X-3 over a wide frequency range obtained using RXTE-ASM and
RXTE-PCA at three observation periods.
}
\label{fig5}
\end{figure}

\subsection{Power Density Spectrum}
The X-ray variation of Cygnus X-3 is dominated by the binary modulation and a
detailed study of its variability characteristics as a function of frequency and
comparison with other black hole candidate sources has not been attempted so far.
van der Klis et al. (\cite{van85}) report the detection of very low frequency regular
oscillations during the rising phase of the binary and Rao et al. (\cite{rao91})
report regular oscillations in hard X-rays. Here we present the Power Density
Spectrum (PDS) for the hard states of Cygnus X-3 with emphasis on the duration of the
multi wavelength monitoring of Choudhury et al. (\cite{cho02a}), after correcting for
the binary variability.

To demonstrate the effect of binary correction on the PDS of this source, we again
consider the longest continuous stretch of pointed observation of this source
shown in Figure \ref{fig3} (Observ. Id. 10126-01-01-020 \& 10126-01-01-02, MJD
50322).
Figure \ref{fig4} shows the PDS after the binary correction. To explicitly
demonstrate that the shape of  residual variability is independent of binary phase,
we show in Figure \ref{fig4} the PDS's separately for the falling part of the light
curve (phase 0.75--1.0 - top panel), rising part of the light curve (phase 0--0.4 -
middle panel), and the peak of the light curve (phase 0.4--0.75 - bottom panel).
Apart from the larger variability due to flares in the rising part (manifest as
increased power below 10$^{-3}$ Hz), the shape of the PDS are similar for all the
three cases. The power-law index in the frequency range 10$^{-3}$ Hz -- 0.1 Hz is
consistent with -1.5 and the total rms power is a few per cent. One of the remarkable
feature of the light curve is the negligible power above 0.1 Hz, and this feature is
observable in the PDS obtained from all the lightcurves, spanning both hard and soft
states (Figure \ref{fig5}). In fact, as shown in Figure \ref{fig5}, the PDS show the
same characteristic featureless powerlaw behaviour in all the states.
 
\subsection{PDS at low frequencies}
Reig et al. (\cite{rei02}) have examined the aperiodic variability of two
micro-quasars Cyg X-1 and GRS 1915+105 at very low frequencies using the RXTE-ASM
dwell data. At frequencies below  10$^{-5}$ Hz it was found that the PDS is
consistent with an index of $\sim$-1 and the rms power below 10$^{-5}$  Hz is 21 --
27 \%, for these two sources. Recently we had argued that Cyg X-3 is an archetypical
black hole binary and have compared its X-ray/ radio emission properties with the
other known black hole binaries (Choudhury et al.\cite{cho03}) and hence it is
instructive to compare the PDS of Cyg X-3 at very low frequencies.

We have derived the PDS of Cyg X-3 at very low frequencies using RXTE-ASM data using
a method similar to that followed by Reig et al.(\cite{rei02}). The results are given
in Figure \ref{fig5}, along with the PDS obtained using the pointed observations of
RXTE-PCA. It is found that below 5 $\times$ 10$^{-7}$ Hz, the power-law index is
flatter (0.97$\pm$0.17), consistent with that obtained for Cyg X-1 and GRS 1915+105.
The remarkable feature of Cyg X-3, however, is the low power - about 3.5\% - in this
frequency range. The flatter power law at low frequencies ($<$ 10$^{-6}$ Hz) and the
steeper power-law at higher frequencies (($>$ 10$^{-3}$ Hz) intercept at
$\sim$10$^{-4}$ Hz signifying a break in the power spectrum. Though there is a hump
like feature at this frequency, we cannot rule out the influence of the binary period
($\sim$10$^{-4}$ Hz) on the shape of this hump. A long uninterrupted observation
would be required to address this issue.

\section {Radio X-ray correlation in Cygnus X-3}
\begin{figure}[t]
\includegraphics{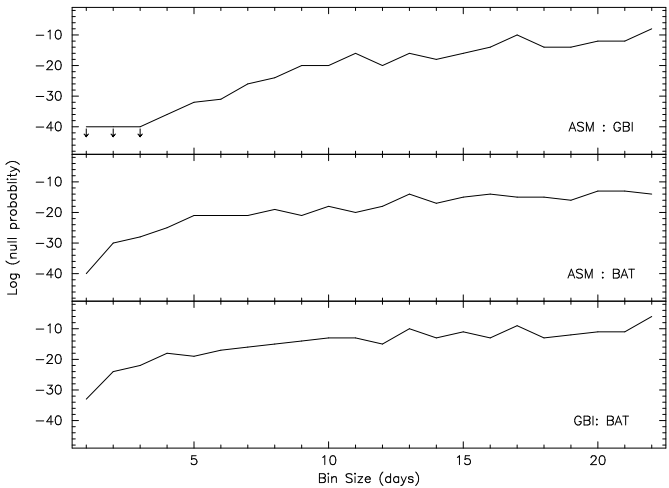}
%\resizebox{\hsize}{!}{\includegraphics[]{grs1915_lc.ps}}
\vskip 6.4cm
\caption{
The null-hypothesis probability for not having a correlation between the soft X-ray
(ASM), radio (GBI) and hard X-ray (BATSE) emissions in Cygnus X-3 for various bin
sizes of integration. The first three points in the top panel denote the upper
limits.}
\label{fig6}
\end{figure}

To avoid the possible effects of binary modulation, the analysis for deriving the
interesting correlation between the X-ray and radio emission in Cygnus X-3, as
reported by Choudhury et al. (\cite{cho02a}), was done using data averaged over 10
days. Here we repeat the same analysis for the binary corrected X-ray emission. The
Spearman partial rank correlation test is used to determine the correlation between
two or more variables. The partial rank coefficient is computed from the sampling
distribution which may be derived by analogy with a parametric statistic (Macklin
\cite{mac82}). For correlation among three variables, say A, X \& Y, the
null-hypothesis is that the correlation between A and X arises entirely from those
of Y with A and X separately. The value of the correlation coefficient lies between
-1 and 1. The negative value signifies anti-correlation. The significance level
associated with the correlation between A and X, independent of Y, is given by the
D-parameter, which is normally distributed about zero with unit variance if the
null-hypothesis, that the A-X relation arises entirely from those of Y with A and X 
separately, is true. The correlation coefficient and other parameters (see Choudhury 
et al. \cite{cho02a,cho03}) are given in table 2 for integration time of data points
ranging from 1 to 15 days. The null-hypothesis probability, as a function of bin
size, is plotted in Figure \ref{fig6} where it appears that the correlation becomes
stronger for smaller bin sizes due to increase in the number of degrees of freedom.
Hence, evidently, both from table 2 and Figure \ref{fig6}, the correlation time scale
is shorter than one day, which is not surprising because for a binary period of 4.8
hours the possible time scales which can come into play, viz. the viscous time scale
of the accretion disc, the variability time scale of the accretion disc corona, and
the time scale for variation of the jet emission must be smaller than a day. 

\begin{table*}
\caption{The Spearman Rank Correlation  coefficient,  the null-hypothesis probability
for no correlation and the D-parameter for the effect of the third parameter, among
the radio (GBI), soft X-ray (ASM) and hard X-ray (BAT) emission of Cygnus X-3, in the
low-hard state with the observed flux averaged for various bin sizes. }
\begin{tabular}{ccccccccccc}
\hline \hline
 Bin Size & No. of & \multicolumn{3}{c}{Spearman Rank Corr. Coeff. } 
  & \multicolumn{3}{c}{Null-hypothesis Probability} 
  & \multicolumn{3}{c}{D Parameter } \\
  (days) & points & ASM:GBI & ASM:BAT &GBI:BAT & ASM:GBI & ASM:BAT &GBI:BAT & ASM:GBI
& ASM:BAT &GBI:BAT \\
\hline
  1 & 638 &   0.70 &  -0.50 &  -0.45 & $<$.10E-40 & .28E-40 & .68E-33 &  17.94 & 
-7.31 &  -4.37 \\
  2 & 356 &   0.73 &  -0.57 &  -0.51 & $<$ .10E-40 & .15E-30 & .49E-24 &  13.83 & 
-6.38 &  -3.22 \\
  3 & 246 &   0.76 &  -0.64 &  -0.58 &  $<$.10E-40 & .13E-28 & .12E-22 &  11.15 & 
-6.11 &  -3.08 \\
  4 & 186 &   0.77 &  -0.68 &  -0.60 & .36E-36 & .12E-25 & .27E-18 &   9.62 &  -6.26
&  -2.17 \\
  5 & 149 &   0.79 &  -0.69 &  -0.66 & .28E-32 & .23E-21 & .79E-19 &   8.73 &  -4.64
&  -3.11 \\
  6 & 125 &   0.82 &  -0.73 &  -0.68 & .49E-31 & .41E-21 & .20E-17 &   8.54 &  -4.75
&  -2.32 \\
  7 & 108 &   0.82 &  -0.77 &  -0.70 & .50E-26 & .19E-21 & .25E-16 &   7.18 &  -5.32
&  -2.06 \\
  8 &  94 &   0.83 &  -0.77 &  -0.72 & .63E-24 & .62E-19 & .17E-15 &   6.80 &  -4.60
&  -2.25 \\
  9 &  84 &   0.81 &  -0.83 &  -0.74 & .95E-20 & .16E-21 & .15E-14 &   5.27 &  -6.05
&  -1.74 \\
 10 &  75 &   0.84 &  -0.82 &  -0.73 & .26E-20 & .56E-18 & .99E-13 &   6.17 &  -5.12
&  -1.19 \\
 11 &  69 &   0.81 &  -0.86 &  -0.75 & .20E-16 & .19E-20 & .76E-13 &   4.33 &  -6.26
&  -1.46 \\
 12 &  63 &   0.87 &  -0.86 &  -0.82 & .88E-20 & .33E-18 & .19E-15 &   5.09 &  -4.25
&  -2.26 \\
 13 &  58 &   0.85 &  -0.81 &  -0.74 & .24E-16 & .74E-14 & .31E-10 &   5.54 &  -4.26
&  -1.13 \\
 14 &  54 &   0.89 &  -0.87 &  -0.82 & .63E-18 & .70E-17 & .19E-13 &   4.90 &  -4.33
&  -1.61 \\
\hline
\hline
\end{tabular}
\end{table*}

\section{Discussion}

\subsection{The binary template and its stability}
A noteworthy feature of Cygnus X-3 is that the binary period has remained consistent
for more than 25 years with a linear decay with no second derivative of the period
(Singh et al. \cite{sin02}). The relative phasing of the X-ray and infra-red
emissions from this source, coupled by the orbital evolution of the line shift of the
He I emission lines, gives the picture that the binary modulation is due to the
orbital motion of the ionized two temperature wind originating in the companion
Wolf-Rayet star irradiated by the X-rays from the compact object (van Kerkwijk
\cite{van93}). The minimum occurs when the cooler part of the wind in the shadow of
the Wolf-Rayet companion is in the line of sight of the observer with the compact
object at the superior conjunction. The asymmetry in the binary modulation profile
may be due to 1) eccentricity in the binary orbit, and/or 2) asymmetric distribution
of matter within the system (Elsner et al. \cite{els80}).
Given the fact that both the binary period and the template have remained consistent
for a period of $\sim$ 25 years, the presence of any reasonable apsidal motion (Ghosh
et al. \cite{gho81}) may be ruled out (van der Klis \& Bonnet-Bidaud \cite{van89}).
Hence, eccentricity, if present in the binary orbit, is negligible to effect any
asymmetry in the lightcurve. Therefore it is more likely that an asymmetric
distribution of matter, causing an asymmetric distribution of optical depth,
introduces the asymmetry in the binary template. In this scenario the orbital period
decay is explained by the loss of angular momentum via the wind from the Wolf-Rayet
companion.

Since the early days of its observation, it is believed that in this source the
X-ray emission is extensively reprocessed, either in the stellar wind (Davidsen \&
Ostriker \cite{dav74}; Hertz et al. \cite{her78}; Becker et al. \cite{bec78}) or in
a cocoon (X-ray halo) surrounding and extending beyond the binary system (Milgrom 
\cite{mil76}, Predehl et al. \cite{pre00}). There also exist models of X-ray
reprocessing in the accretion disk corona (White \& Holt \cite{whi82}; Molnar \&
Mauche \cite{mol86}) where the low energy photons are upscattered by Comptonization
also producing fluorescence of ionized Fe (Rajeev et al. \cite{raj94}). Nakamura et
al. (\cite{nak93}) explain the X-ray spectra as obtained by the Ginga observatory by
proposing the presence of three species of ionized gas, fully ionized, almost fully
ionized and nearly ionized, engulfing the binary system. Fender et al. (\cite{fen99})
suggest a WR type wind with the geometry of a disc in the binary plane with a size
much bigger than the binary system to be the origin of the binary modulated He
emission lines (obtained from infra-red spectra). This wind originates from the
companion which is a WN type Wolf-Rayet star. Polarimetric study of the K-band
lightcurve also suggests a preferential plane of scattering. In this model the X-ray
emission undergoes scattering in this disc-like two temperature wind (van Kerkwijk et
al. \cite{van93}), resulting in the asymmetry in the binary template.

The residue of the (binary) folded lightcurve (Figure \ref{fig2}) may be attributed
to the long term variation of the X-ray emission by virtue of the change in wind
and/or cocoon mass distribution, including various state changes from soft (and high)
to low (and hard) and vice-versa, accompanied by the correlated radio flares. One
small aspect not considered so far is the generally random fluctuation in the X-ray
lightcurve at the peak of the binary ephemeris. A very detailed analysis of the
nature of this fluctuation may help in determining the geometrical and physical
structure of the accreting system involving the wind from the companion and the X-ray
halo (Predehl et al. \cite{pre00}) engulfing the system. Recently, Stark  \& Saia
(\cite{sta03}) have attempted to constrain the binary orbital parameters by measuring
the line shift of Helium $\alpha$ line of Fe XXV and Lyman $\alpha$ lines of Si XIV
and S XVI as a function of the orbital phase. By ascribing the Fe line to originate
very close to the compact object surface and the Si and S line to originate from the
wind from the companion, they constrain the masses of the companion and the compact
object to be $\le 7.3 M\odot$ and $\le 3.6 M_\odot$, respectively, provided the
orbital inclination is small ($i=24^\circ$ - this agrees with the limit $0-40^\circ$
of Fender et al. \cite{fen99}). The phase resolved spectra of the various Fe emission
lines (6.4 keV, 6.7 keV \& 6.9 keV) and the absorption edges (7.1 keV \& 9.1 keV)
during the both low-hard and high-soft states may provide better constraints on the
origin of the Fe emission lines. The binary modulation of the emission lines reported
by Stark and Saia (\cite{sta03}) need to be fit properly with the binary template
(Singh et al. \cite{sin02}) to obtain a more definitive mass function of the system.
A long continuous X-ray observation of the source spanning a few orbital periods with
very high resolution X-ray spectra will provide a better binary profile which is
essential for constraining the binary parameters and obtaining the geometrical and
physical structure of the system.

\subsection{Power Density Spectrum}
The distinct feature of the power density spectra of the source is the shifting of
the spectra towards low frequency regime (vis-a-vis the `normal' frequency regime of
other Galactic X-ray binaries), corresponding to very massive black hole systems
(Hayashida et al. \cite{hay98}, Czerny et al. \cite{cze01}). Sunyaev \& Revnivtsev
(\cite{sun00}) have compiled the power density spectra (multiplied by frequency) of
the most common Galactic X-ray binaries, both neutron stars and black hole candidates
(in their low/hard states). The black hole binaries show, typically, a power law
dependence with a positive index in the region of $\sim$0.01 -- 1 Hz, flat spectra
for the next decade of frequency range, followed by a power law decay (i.e. negative
index) of power in the $\sim$10 --100 Hz. The PDS of the neutron stars is generally
shifted towards the higher frequency region by an order of magnitude. From the PDS of
Cygnus X-3 (Figure \ref{fig4}) it is quite apparent that, from the RXTE--PCA pointing
observations, we are catching the power law decay portion of the PDS, shifted by a
few decades in the lower frequency regime, with the power getting merged with the
white noise at $\sim$ 0.1 Hz. Due to improved continuous observational capabilities
of RXTE it is possible to extend the PDS to the lower frequency regime
($\sim 10^{-4}$ Hz) and hence ascertain from the wide frequency band PDS that, in
general, the power above $\sim$ 0.1 Hz is not discernible from the white noise. One
may reconcile the absence of power in the high frequency regime to the scattering of
the X-ray photons in the wind from the companion Wolf-Rayet star, reducing the
amplitude of the fast X-ray variability (Berger \& van der Klis \cite{ber94}). If we
reconcile the absence of power above 0.1 Hz to the reprocessing of the X-ray emission
in the circumstellar environment, then this feature does provide an interesting
sidelight to the paradigm which states that the variability time-scale scales
linearly with the mass of the compact object (Hayashida et al.\cite{hay98}, Czerny et
al.\cite{cze01}), and introduce an additional factor, viz. reprocessing in the dense
circumstellar material, where it exists, into the picture. It is also interesting
that the PDS at very low frequencies have about an order of  magnitude lower power
compared to other black hole candidate sources Cyg X-1 and GRS 1915+105.

\subsection{X-ray radio correlations}
Correlation between soft X-ray and radio emission has been noted in many black hole
binary sources (Gallo et al. \cite{gal02}) and finding similar correlation in Cygnus
X-3 (Choudhury et al. \cite{cho02a}) firmly puts Cygnus X-3 as a good candidate for
harboring a black hole. Choudhury et al. (\cite{cho03}) point out that correlation
between the X-ray and radio emission and the anti-correlation between the soft and
hard X-rays are directly related to a causal connection between the spectral shape in
the X-rays and the radio emission. They have shown that such a picture consistently
explains the observed behaviors of three bright Galactic X-ray binary sources Cygnus
X-3, Cygnus X-1 and GRS 1915+105. By correcting the X-ray light curve for the binary
variations, we have shown that the correlation time scale is shorter than a day
(Table 1 \& Figure \ref{fig6}). Therefore it is necessary to analyze the pointed
continuous observation in the X-ray wide-band in order to determine the dynamic time
scale of soft and hard X-ray emission. This will enable us to understand the detailed
structure of the accretion disk, and is the focus of our future work. A simultaneous
radio observation will be imperative in understanding the detailed mechanism of the
disk jet connection in the system, which will provide a quantitative estimate of the
extent of jet power being emitted in the X-ray band.

\section{Conclusions}
In this paper we have suggested a method to correct the  X-ray light curve of Cygnus
X-3 for the binary modulation. The binary period shows consistency over a long period
of 25 years, with minor variation in the asymmetric binary phase profile. An X-ray
halo consisting of photoionized plasma originating from the wind of the companion
Wolf-Rayet type companion star reprocesses the X-ray emission, while any eccentricity
of the binary orbit which may cause asymmetry in the X-ray binary profile is minimal
because of absence of any detectable apsidal motion over the long period of $\sim$ 25
years.

We have also examined the binary corrected light curve for short term variabilities 
and noted that the shape of the PDS follows the power law behavior in both low and
high states, similar to other Galactic black hole sources. Most significantly there
exists no power at frequency $>$ 0.1 Hz, with the PDS being shifted to the low
frequency regime vis-a-vis the Galactic X-ray binaries, both black hole and neutron
stars. This is a unique feature of this enigmatic source.

Using the corrected light curve we have examined the X-ray radio correlations at
various time scales and have concluded that the time scale is shorter than a day. The
soft and hard X-ray correlation need to be examined using the binary corrected RXTE
pointed (PCA and HEXTE) data covering a time scale of a few hours or longer to
unambiguously determine the time scale of anti-correlation, and hence determine the
detailed physical as well geometrical structure of the disk.

\section*{Acknowledgments} This research has made use of data obtained 
through the HEASARC Online Service, provided by the NASA/GSFC, and the Green Bank
Interferometer, a facility of the National Science Foundation operated by the NRAO in
support of NASA High Energy Astrophysics Programs. MC and SVV have been partially
supported by the Kanwal Rekhi Scholarship for Career Development. AKJ is grateful to
P.~S.~Goel, Director, ISAC and K.~Kasturirangan, Chairman, ISRO, for their constant
encouragement and support during the course of this work. The authors acknowledge the
editor A. Jones for his continued patient attention, and the referee M. Stark for
very detailed assessment and extremely critical revision of the paper.

\end{document}